\documentclass[universe,article,accept,pdftex,moreauthors]{Definitions/mdpi} 

\firstpage{1} 
\pubvolume{1}
\issuenum{1}
\articlenumber{0}
\pubyear{2024}
\copyrightyear{2024}
\externaleditor{\textls[-25]{Academic Editor: Firstname Lastname}}
\datereceived{20 May 2024} 
\daterevised{14 July 2024} 
\dateaccepted{16 July 2024} 
\datepublished{ } 
\hreflink{https://doi.org/} 

\Title{Study of Wide-Field-of-View X-ray Observations of the Virgo Cluster Using the Lobster Eye Imager for Astronomy
} 

\TitleCitation{Study of Wide-Field-of-View X-ray Observations of the Virgo Cluster Using the Lobster Eye Imager for Astronomy}

\Author{Wen-Cheng Feng $^{1, 2}$\orcidA{}, Shu-Mei Jia $^{2,}$\orcidB{}*, Hai-Hui Zhao $^{1,}$*\orcidC{}, Heng Yu $^{3}$\orcidD{}, Hai-Wu Pan $^{4}$, Cheng-Kui Li $^{2}$\orcidE{}, Yu-Lin~Cheng $^{5}$\orcidG{}, Shan-Shan Weng $^{1}$\orcidF{}, Yong Chen $^{2}$, Yuan Liu $^{4}$, Zhi-Xing Ling $^{4}$ and Chen Zhang $^{4}$}

\AuthorNames{Wen-Cheng Feng, Shu-Mei Jia, Hai-Hui Zhao, Heng Yu, Cheng-Kui Li, Yu-Lin Cheng, Shan-Shan Weng, Yong Chen, Yuan Liu, Zhi-Xing Ling, and Chen Zhang}

\AuthorCitation{Feng, W.-C. et al. }

\address{%
$^{1}$ \quad Department of Physics and Institute of Theoretical Physics, Nanjing Normal University, \linebreak{}Nanjing~210023, China; fengwc@ihep.ac.cn (W.-C.F.)\\  
$^{2}$ \quad  Key Laboratory of Particle Astrophysics, Institute of High Energy Physics, Chinese Academy of Sciences, Beijing~100049, China\\
$^{3}$ \quad  Department of Astronomy, Beijing Normal University, Beijing~100875, China; yuheng@bnu.edu.cn\\
$^{4}$ \quad  National Astronomical Observatories, Chinese Academy of Sciences, Beijing~100101, China\\
$^{5}$ \quad  Department of Astronomy, Yunnan University, Kunming~650091, China}

\corres{Correspondence: jiasm@ihep.ac.cn (S.-M.J.); zhaohh@njnu.edu.cn (H.-H.Z.)}

\abstract{The Lobster Eye Imager for Astronomy (LEIA) is the pathﬁnder of the wide-ﬁeld X-ray telescope used in the Einstein Probe mission. In this study, we present an image of the Virgo Cluster taken by LEIA in the 0.5--4.5 keV band with an exposure time of $\sim$17.3 ks in the central region. This extended emission is generally consistent with the results obtained by ROSAT. However, the field is affected by bright point sources due to the instrument's Point Spread Function (PSF) effect. Through fitting of the LEIA spectrum of the Virgo Cluster, we obtained a temperature of $2.1^{+0.3}_{-0.1}$ keV, which is consistent with the XMM-Newton results ($\sim$2.3 keV). Above 1.6 keV, the spectrum is dominated by the X-ray background.
In summary, this study validates LEIA's extended source imaging and spectral resolution capabilities for the first time.
}


\begin{document}

\section{Introduction}
\label{introduction}

Lobster-eye micropore optics (MPO) is an novel X-ray focusing technology that is known for its ability to observe a wide field of view and its true imaging capabilities. Since Angel \cite{1979ApJ...233..364A} first introduced a design of an X-ray all-sky monitor (ASM) based on lobster-eye MPO, several teams have conducted extensive research on lobster-eye optics  \cite{1989RScI...60.1026W,1991RScI...62.1542C,10.1117/12.51224,1993NIMPA.324..404F,10.1117/12.2055434,2017SPIE10567E..19H} and have developed conceptual designs for lobster-eye X-ray ASMs  \cite{10.1093/mnras/279.3.733,10.1117/12.454217}. Real missions using lobster-eye ASMs have been proposed \cite{2016SSRv..202..235Y,10.1117/12.2561301}. The first instrument constructed for a formal mission was the MIXS \cite{2020SSRv..216..126B} aboard BepiColombo, which consists of a $1^{\circ}$ field-of-view (FoV) Wolter telescope and a $10^{\circ}$ FoV collimator, both constructed using MPO. Its debut observations are expected within a few years, when the mission reaches Mercury. Several X-ray telescopes based on lobster-eye MPO are currently in development and scheduled for launch in the coming years, including SVOM-MXT \cite{10.1117/12.2232484,10.1117/12.2628635} and SMILE-SXI \cite{2016AGUFMSM44A..04S}.

The Lobster Eye Imager for Astronomy (LEIA), launched in July 2022 onboard the SATech-01 satellite, is the first lobster-eye wide-field focusing X-ray telescope \citep{Ling_2023}. It features a mostly unvignetted FoV measuring $18.6^{\circ} \times 18.6^{\circ}$ in the energy band 0.5--4.5keV. It serves as a pathfinder for the wide-field X-ray telescope in the Einstein Probe mission and was designed to verify the in-orbit performance of the EP-WXT and optimize its instrumental parameters and operational conditions. LEIA achieves a spatial resolution ranging from $4^{\prime}$ to $7^{\prime}$ of the full width at half maximum (FWHM), and its effective area of 2--3 cm$^2$ shows only mild variations across almost the entire FoV \citep{Zhang+etal+2022}.

LEIA has been operating in orbit for more than one year, with Virgo being the first galaxy cluster observed. The Virgo Cluster is the closest ($\sim$16.1 Mpc \cite{2001ApJ...546..681T}) bright irregular cluster \citep{Bohringer+etal+1994}. The brightest cluster galaxy is the giant elliptical galaxy M87 (NGC 4486), which contains a super-massive black hole. The Virgo Cluster was first captured by the ROSAT all-sky survey \citep{Truemper+1992} over the course of three months from November 1990 to January 1991. With an average exposure time of $\sim$450 ks, the X-ray image of the Virgo Cluster is clear at the 0.1 to 2.4 keV band observed by ROSAT. The large-scale structure of the Virgo Cluster was first studied in detail by \citet{Bohringer+etal+1994}, and the results revealed  an extended emission in the $12.8^{\circ} \times 12.8^{\circ}$ area of the Virgo region, with most of the cluster's mass being concentrated in galaxy M87. Using ASCA, \citet{Shibata+etal+2001} was the first to derive a full temperature map of the Virgo Cluster, covering an area of 19 deg$^2$. \linebreak{}\citet{Urban+etal+2011} analyzed a mosaic comprising thirteen XMM-Newton images observed from the center of the cluster to the north with a radius of approximately 1.2 Mpc and obtained a mean temperature of $\sim$2.3 keV.

In this study, we present an X-ray image of the Virgo Cluster as observed by LEIA. The image is affected by LEIA's PSF, but the main radiation results are consistent with the results obtained from ROSAT. We also analyzed the spectra taken on the Virgo Cluster and obtained temperature fitting results consistent with those obtained by XMM-Newton. This work aims to validate LEIA's capabilities in the study of extended sources.
The paper is arranged as follows: in Section \ref{data}, we describe the observation data used. The pictures of the Virgo Cluster and the spectrum fitting are described in Section \ref{results}. The Discussion and Conclusions are given in Section \ref{discussion} and Section \ref{conclusions}, respectively. The redshift (z) adopted for M87 is 0.00428. In this work, $H_0 = 67.8 \ \rm km \ s^{-1} \  Mpc$, $\Omega_m=0.308$, and $\Omega_{\Lambda}=0.692$ \citep{Cappellari+etal+2011}.

\section{Observations and Data Reduction}
\label{data}
\subsection*{LEIA Observations}
A total of 36 observations of the Virgo Cluster were used. The observations were carried out between 19 December 2022  and  26 February 2023. The telemetry data from LEIA were downloaded to the ground station each day and sent to the data center at the National Space Science Center (CAS), where the telemetry data were decoded, unpacked, verified, and processed to remove duplicates. Then, the data were sent to the science center at the National Astronomical Observatories (CAS), where the data were converted into standard FITS format and processed by a software pipeline and the corresponding calibration database (CALDB) to calibrate the events and extract high-level products in the following steps \cite{Zhang+etal+2022,2024arXiv240416425L}:

\begin{enumerate}
  \item Since the digital number (DN) of each event was only subtracted by the corresponding row's median value on-board, the ground processing pipeline further subtracted the bias residual of each pixel stored in CALDB.
  \item The position of each event was converted to the celestial coordinates (J2000) using the satellite attitude and alignment matrix between the detector and the satellite. The nonlinear distortion due to the non-spherical shape of MPO was also corrected for in this step according to the result of the on-ground calibration performed at IHEP.
  \item The anomalous pixels stored in CALBD and their nearby pixels were flagged. New anomalous pixels (hot and flickering) were also searched for and flagged.
  \item A grade and a single PHA value were assigned to the event according to the DNs of the $3\times3$ pixels around the event.
  \item The Pulse-Invariant (PI) value of each event was calculated according to the gain stored in CALDB.
  \item Single, double, triple, and quadruple events without any anomalous flag were selected for further processing. The geomagnetic cut-off rigidity (COR) was employed to remove the high background interval in the high-latitude region, i.e., $\rm{COR}>5$. The earth elevation angle was required to be larger than 10 degrees to remove the effect of earth occultation. The angular distance to the nominal pointing should be smaller than 0.2 degrees to ensure the stabilization of the satellite.
  \item The pipeline generated an exposure map that accounts for bad CMOS pixels and columns, attitude variations, and telescope vignetting (optional) for event files.
  \item An image in the 0.5--4.0\,keV band was accumulated for each CMOS using the cleaned events, and source detection was performed on each image to generate a source catalog.
  \item The light curve and spectrum of each source in the catalog were extracted. The pipeline also generated the corresponding Response Matrix File (RMF) and Ancillary Response Function (ARF) file, which accounted for the effective area, vignetting correction, and PSF correction.
\end{enumerate}
	
The above procedures were conducted on each CMOS separately. The resulting event files and images can be further merged to search for faint sources.
 
\section{Results}
\label{results}
\subsection{The X-ray Image \label{sec3.1}}
\begin{figure}[H]
	\begin{adjustwidth}{-\extralength}{0cm}
	\centering
\includegraphics[width=8.19cm, angle=0]{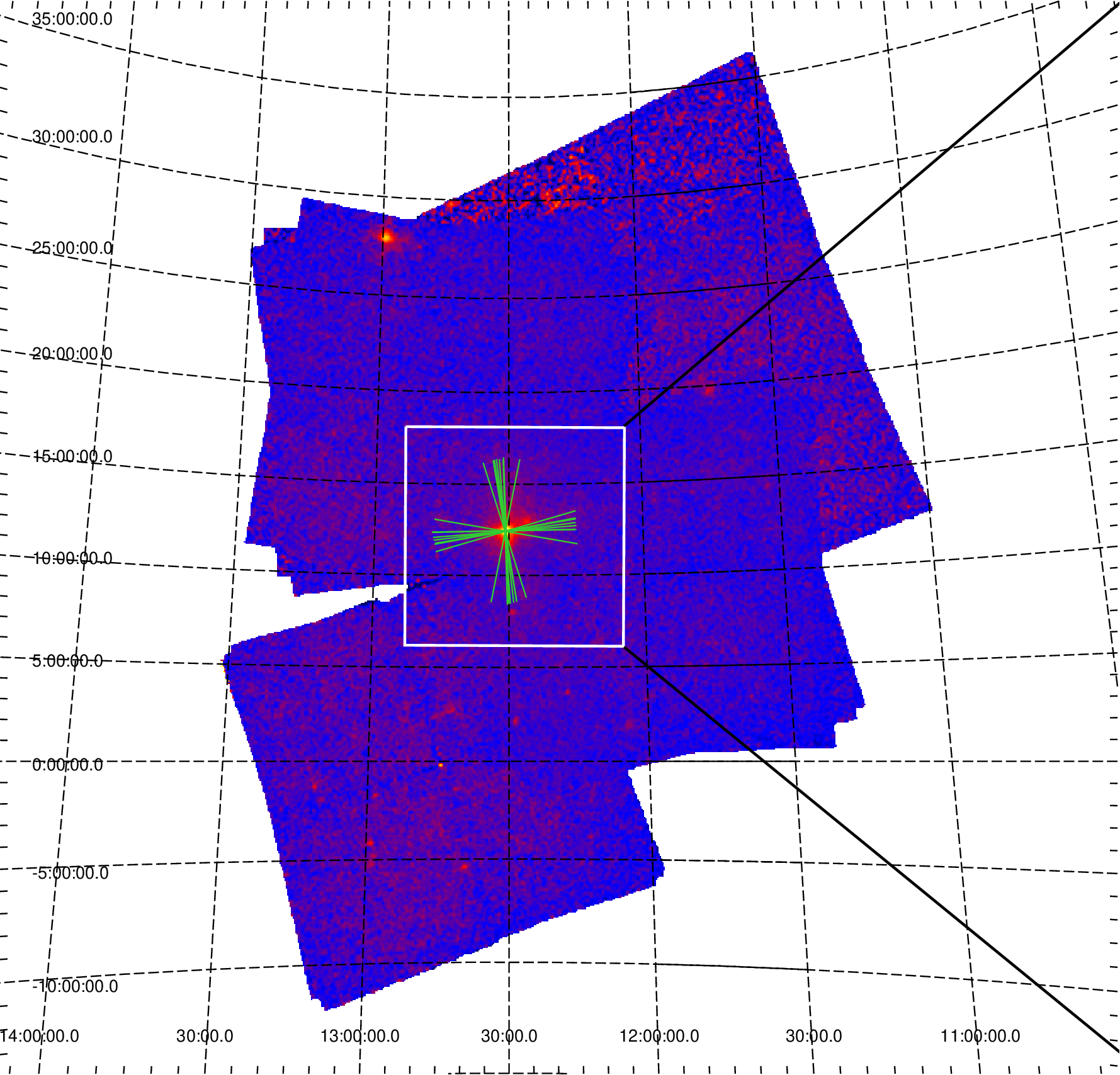}
    \hspace{0.7cm}
    \includegraphics[width=8cm, angle=0]{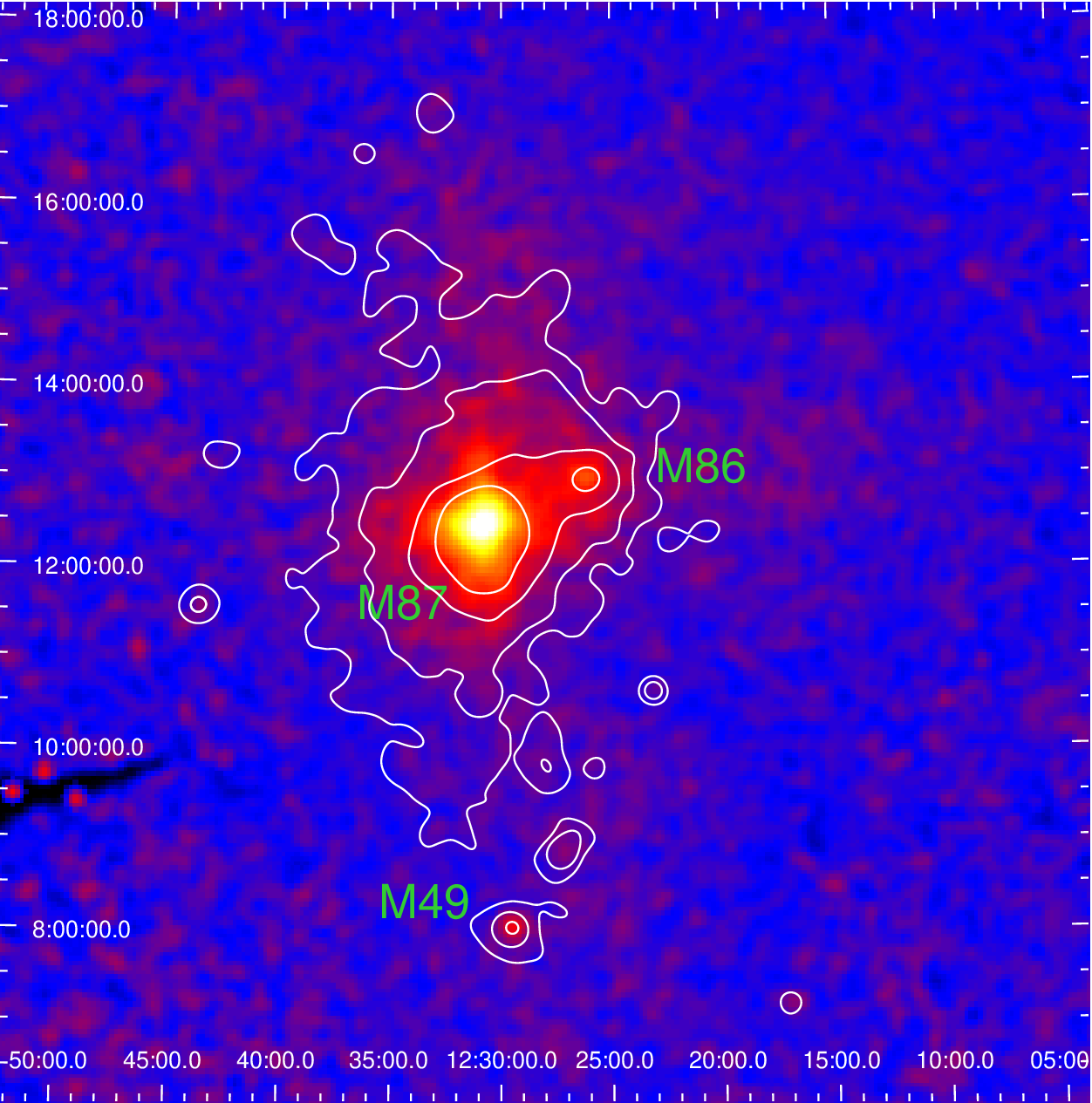}
\end{adjustwidth}
	\caption{ (\textbf{Left}) The exposure-corrected mosaic image of the Virgo Cluster observed by LEIA in the 0.5--4.5 keV band with an exposure time of $\sim$17.8 ks at the center position. Green crosses indicate the direction of each observation. (\textbf{Right})  The image in the right panel is an enlarged view of the white box ($12^{\circ} \times 12^{\circ}$) in the image on the left, which overlaps the contour of ROSAT PSPC image of the Virgo Cluster in the 0.5--2.0 keV band. The contour lines represent being 3$\sigma$, 5$\sigma$, 11$\sigma$, and 25$\sigma$ above the background \citep{Bohringer+etal+1994}.  }
	\label{Fig1}
\end{figure}

We used the Montage Image Mosaic Engine
\endnote{Available from \url{http://montage.ipac.caltech.edu/}} 
 to combine LEIA's observations of the Virgo Cluster and its surroundings to obtain the exposure-corrected mosaic image in the 0.5--4.5~keV band shown in Figure ~\ref{Fig1}. The exposure time of the center position is $\sim$17.8~ks. The green crosses indicate the direction of each observation. The brightest core is the central galaxy M87. Its emissions extend several degrees outward in the shape of a cross. We then created a white box (shown in \mbox{Figure~\ref{Fig1}}) that was $12^{\circ} \times 12^{\circ}$ in size and centered around M87 to frame the Virgo Cluster to further investigate its structure. The right panel of \mbox{Figure ~\ref{Fig1}} is an enlarged image of the box region. The white contours were obtained from the smoothed ROSAT Position-Sensitive Proportional Counter (PSPC) image of the medium X-ray emission of the Virgo Cluster's intracluster in the 0.5--2.0 keV band~\citep{Vollmer+etal+2013}. \linebreak{}The contour lines represent being 3$\sigma$, 5$\sigma$, 11$\sigma$, and 25$\sigma$ above the background \citep{Bohringer+etal+1994}.
The elliptical galaxy M86 \citep{White+etal+1991} lies northwest of M87. Another prominent feature in Figure ~\ref{Fig1} is the X-ray halo around M49 \citep{Binggeli+etal+1987}, which is located south of M87. Some structures that extend beyond the ROSAT contour in the far north and west in the LEIA images are affected by the PSF effect \citep{Zhang+etal+2022} and not the real cluster structure. We provide a detailed discussion on the arm structure in Section \ref{discussion}.

\vspace{3pt}

\begin{figure}[H]
	\begin{adjustwidth}{-\extralength}{0cm}
\centering
	\includegraphics[width=5.7cm, angle=0]{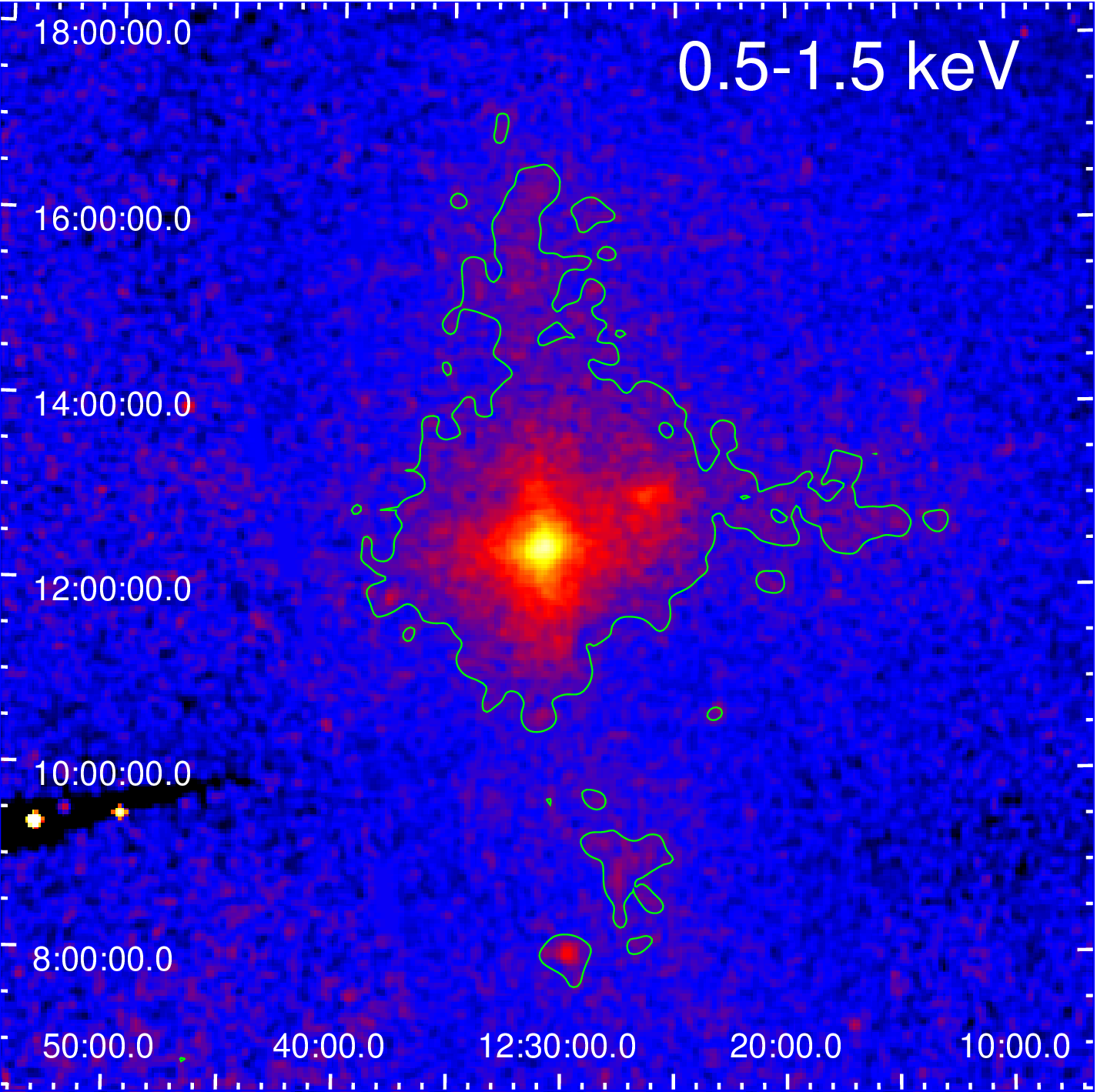}
    \hspace{0.3cm}
    \includegraphics[width=5.7cm, angle=0]{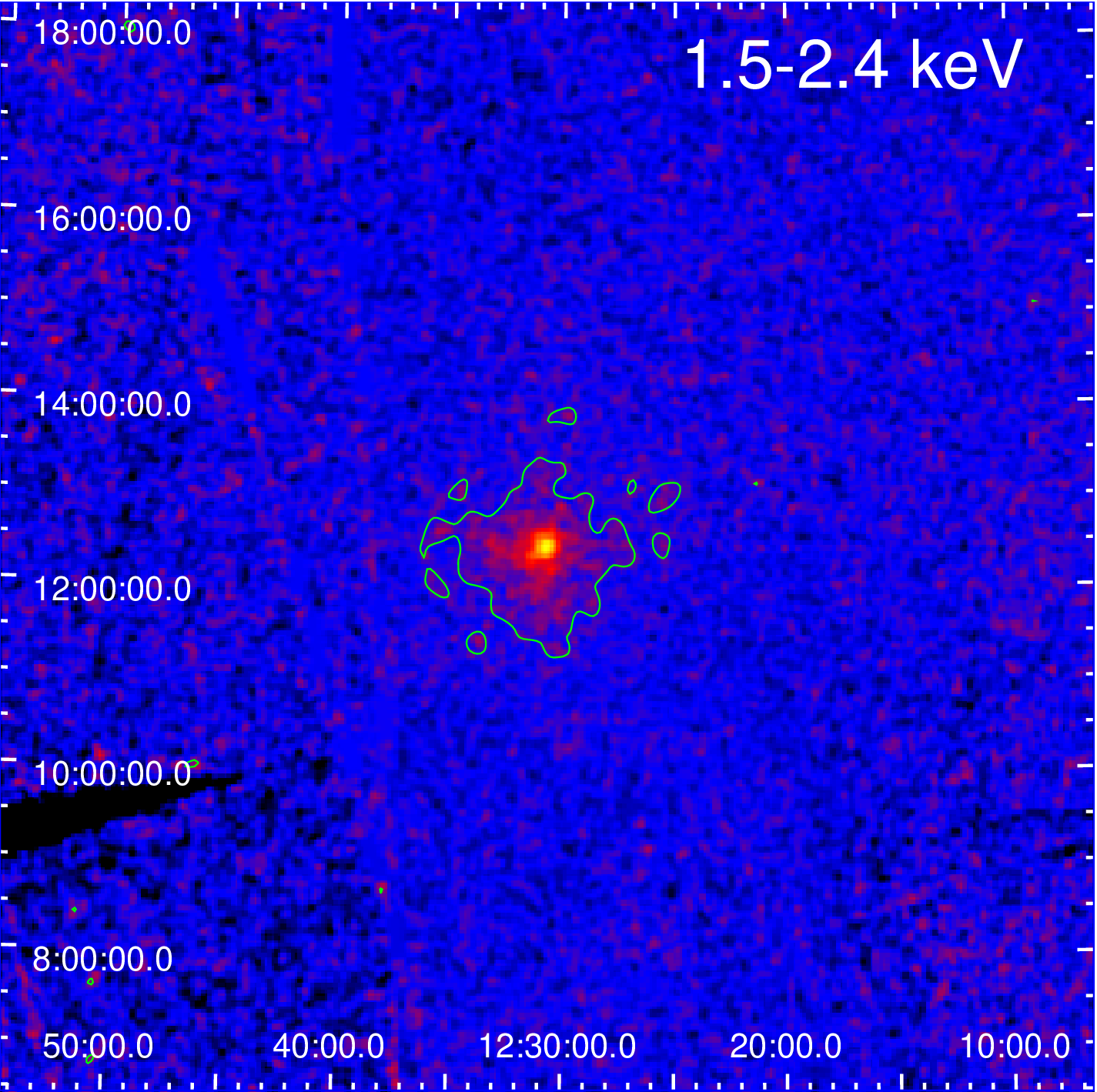}
    \hspace{0.3cm}
    \includegraphics[width=5.7cm, angle=0]{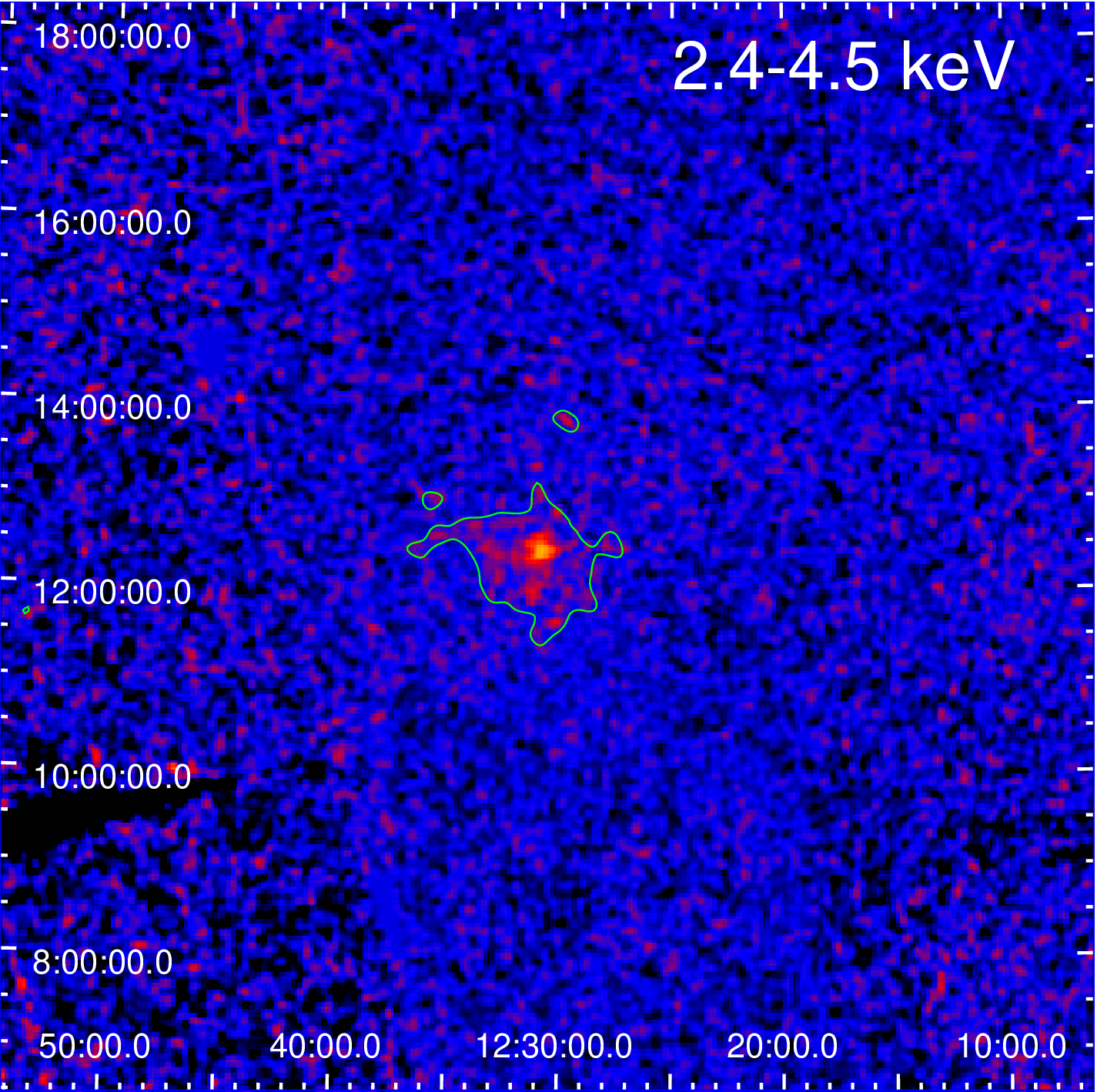}
\end{adjustwidth}
	\caption{ A LEIA image of the Virgo Cluster in different energy bands. From left to right: 0.5--1.5 keV, 1.5--2.4 keV, and 2.4--4.5 keV. For comparison, the contour shows 5$\sigma$ levels above background noise for the image in the 0.5--1.5 keV band and 3$\sigma$ levels above background noise for the image in the 1.5--2.4 keV and 2.4--4.5 keV bands.}
	\label{Fig2}
\end{figure}

As the ROSAT PSPC observations only offer an energy band that extends from 0.1 to 2.4 keV, LEIA's wide-field X-ray imaging up to 4.5 keV provides us with an opportunity to see the features of the Virgo Cluster. 
In order to better reveal the radiation distribution of the Virgo Cluster, Figure ~\ref{Fig2} presents a LEIA image of the source in the emission ranges of 0.5--1.5 keV, 1.5--2.4 keV, and 2.4--4.5 keV from the central 6$^{\circ}$ point. For comparison, we also provide a contour in each panel which shows 5$\sigma$ levels above the background noise for the images in the 0.5--1.5 keV range and 3$\sigma$ levels for the images in the 1.5--2.4 keV and 2.4--4.5 keV bands. In the 0.5--1.5 keV band, the diffuse arm structure around M87 is clearly visible. The scale of arm extension to each edge is up to 5$^{\circ}$; above the 1.5 keV band, the thermal gas radiation is barely visible, overshadowed by the central point source and the influence of its PSF. This is the first time a lobster-eye focusing telescope has been used to see a large-scale image of the Virgo Cluster that shows that diffuse thermal emission exists around M87 in the 0.5--4.5 keV band.

\vspace{3pt}

\subsection{LEIA Spectrum}
\label{LEIA}
\begin{figure}[H]
	\includegraphics[width=9.5cm]{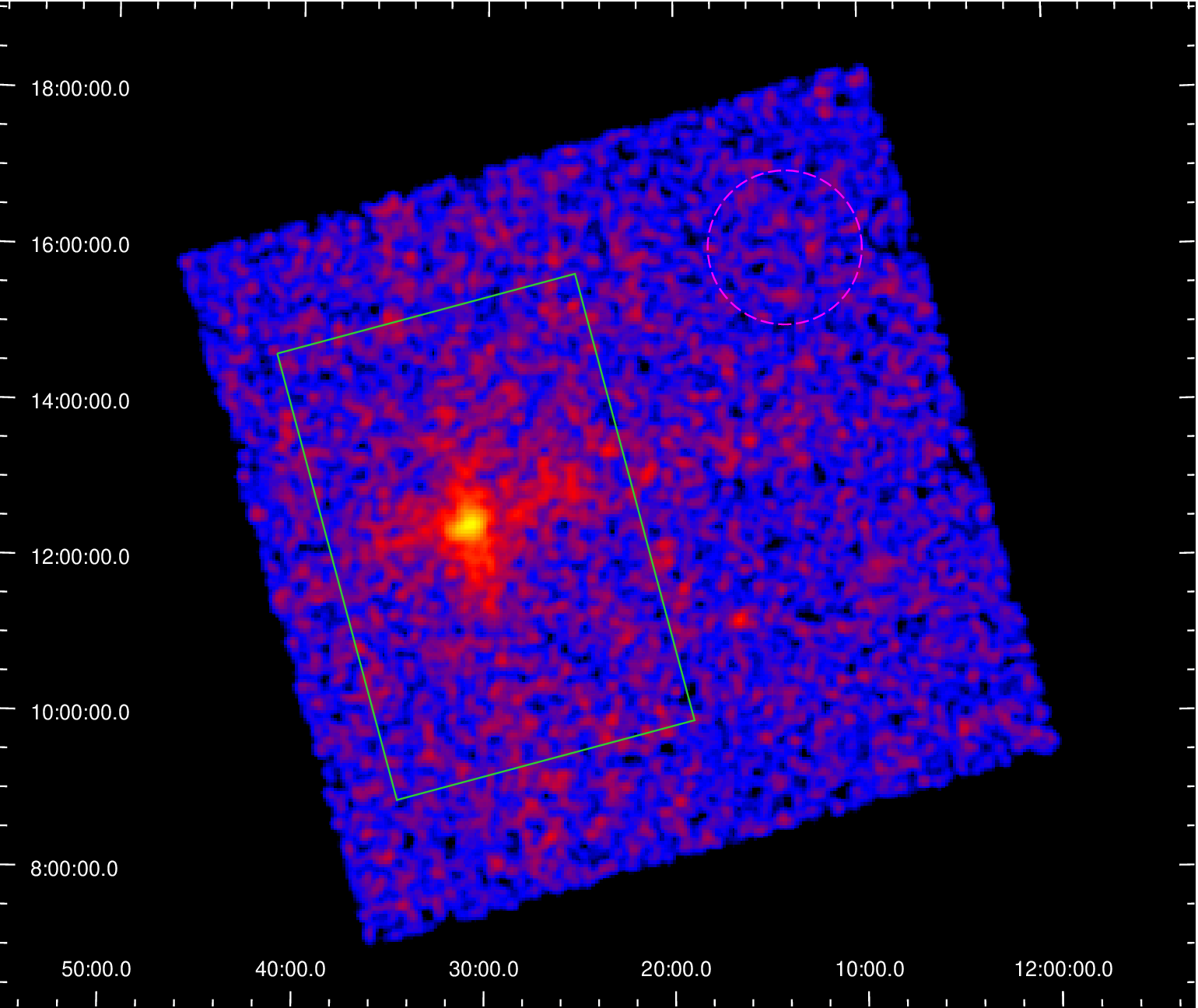}
	\caption{X-ray image of the Virgo Cluster obtained by LEIA during a one-shot observation of 1083 s in 0.5--4.5 keV. The solid-lined green box is the source region of the Virgo Cluster. The magenta circle formed with a dashed line is the background region. }
	\label{Fig3}
\end{figure}

We selected five observations containing the majority of the thermal gas emissions and relatively high-quality data for spectral analysis. To extract the spectra of the Virgo Cluster, we adopted a box region that was $4^{\circ} \times 6^{\circ}$ in size and included M87 (shown in Figure ~\ref{Fig3}, the same size as the source region for every observation). To eliminate the influence of the cosmic background, we circled an area with a $1^{\circ}$ radius and where there were no resolved point sources in the background region. The spectra were extracted in XSELECT (version 2.4). We merged these spectra together using the \texttt{addspec} command to enhance the data statistics. Then, we binned the merged spectra to ensure that 1 bin contained more than 200~photons. 

\begin{figure}[H]
	\includegraphics[width=9.8cm, angle=270]{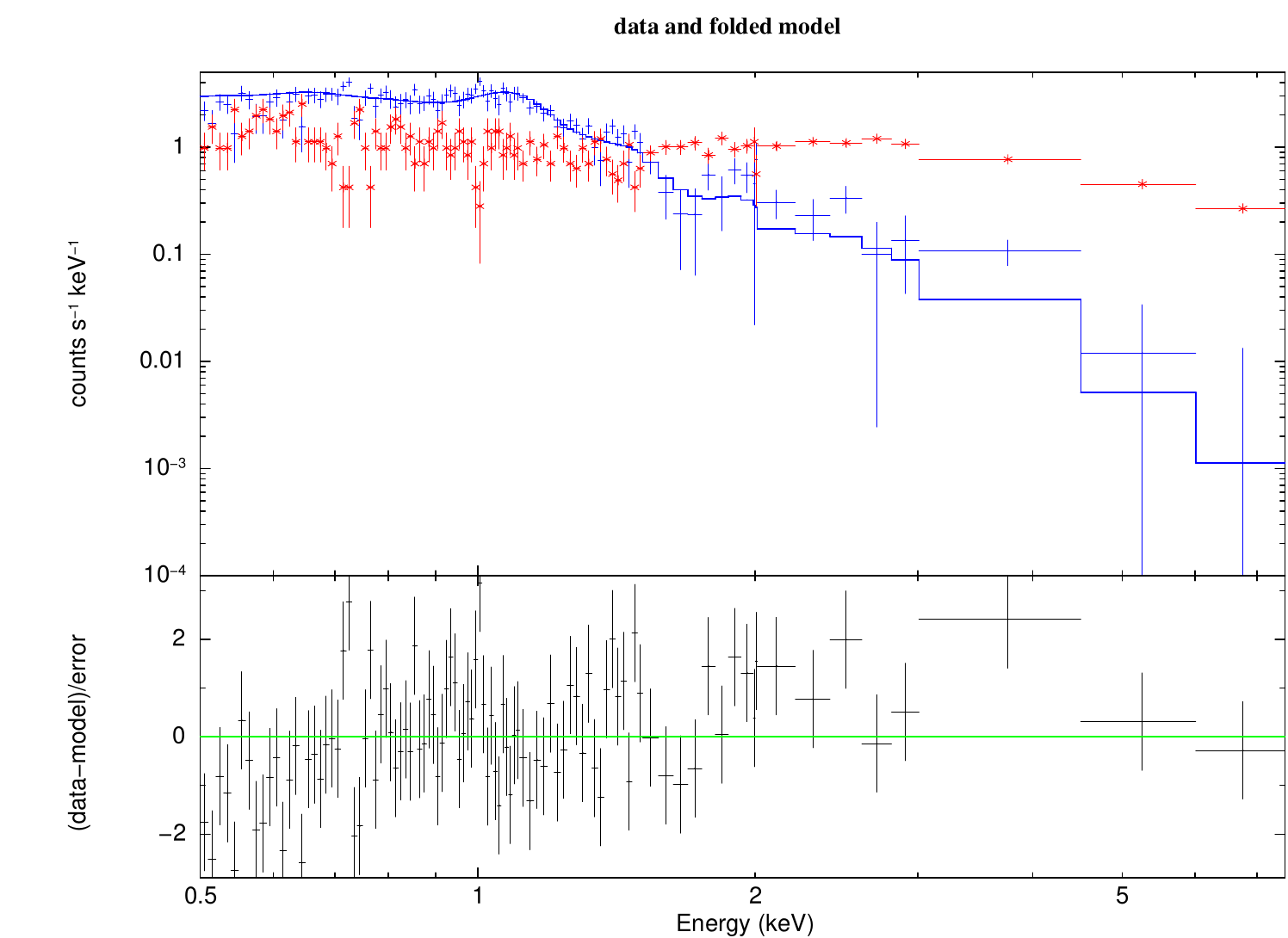}
	\caption{LEIA spectrum of the Virgo Cluster. Blue and red data points are from the source and background regions, respectively. The solid blue line shows the fit of an absorbed APEC model with the abundance fixed to 0.3$Z_{\odot}$.}
	\label{Fig4}
\end{figure}

We used XSPEC (version 12.11.1) to model the spectrum. Most regions of the Virgo Cluster can be well fitted by a single temperature model \citep{Simionescu+etal+2007}, while the central region ($<$2 arcmin) may be affected by non-thermal radiation \cite{Matsushita+etal+2002, Molendi+2002}. The LEIA spectrum can be characterized by an absorbed APEC model, allowing the model and the spectrum to be fit together with the WABS*APEC fitting model. The hydrogen column density was fixed as $2.13 \times 10^{20} \ \rm cm^{-2}$ \cite{Kalberla+etal+2005, Urban+etal+2011}. The redshift was fixed to 0.00428. Due to the inability to determine the metal abundance effectively from the spectra, we set the abundance ratio to 0.3$Z_{\odot}$. 
The fitting results are listed in Table ~\ref{Tab1} and shown in Figure ~\ref{Fig4}. 
The blue and red data points are from the source and background components, respectively.
The energy spectrum above 1.6 keV is dominated by the background caused by the aperture and non-aperture cosmic X-ray background (CXB), as well as the particle-induced background \cite{2017ExA....43..267Z}.
The reduced chi-squares of $\sim$1.24 show that our model is reasonable. 
The temperature of the model fitting is $2.1^{+0.3}_{-0.1}$ keV, which is consistent with the overall temperature ($2.4^{+0.3}_{-0.2}$ keV) within the 1$^{\circ}$ radius of M87 measured by ROSAT \citep{Bohringer+etal+1994} and
the average temperature of $\sim$2.3 keV \citep{Urban+etal+2011} derived by \citet{Arnaud+etal+2005} obtained from XMM-Newton.

\vspace{2pt}
\vspace{-10pt}

\begin{table}[H]
		\caption{The parameters of spectrum fitting of LEIA in the 0.5--4.5 keV band. The uncertainties of the parameters were calculated with a confidence level of 68\%. The fitting model for the spectrum is WABS*APEC.}\label{Tab1}
        \setlength{\tabcolsep}{22pt}
		\newcolumntype{C}{>{\centering\arraybackslash}X}
\begin{tabularx}{\textwidth}{llll}
			\toprule
			\textbf{Model Component }& \textbf{Parameters} &  \textbf{LEIA} \\
			 \midrule
			WABS   &  nH ($10^{20}$ cm$^2$)   & 2.13 (frozen)    \\		
			APEC      &  KT (keV)    &      $2.1^{+0.3}_{-0.1}$  \\		
			           &  Abun (metal abundances)       &  0.30 (frozen)          \\	
			           &  z (redshift)        &  0.00428  (frozen)       \\ 
                       &  norm       &   $0.99^{+0.04}_{-0.02}$        \\
            $\chi^2$/DOF   &                &   187.62/151    \\       
			\bottomrule
		\end{tabularx}
\end{table}

\section{Discussion}
\label{discussion}
The LEIA image is generally consistent with the ROSAT contours, with slight differences in the north and west, as described in Section \ref{sec3.1}. It is necessary to determine whether the extra structures in the LEIA image are the real components of the Virgo Cluster. Due to the design principle of the lobster-eye optical device, LEIA's PSF exhibits a cross shape. LEIA's mirror is composed of numerous small square apertures that are densely arranged on a spherical surface and point towards the common curved center of the sphere. When X-rays enter these apertures, they undergo grazing incident reflection on the inner walls of the apertures, ultimately focusing on the focal plane and forming a cross-shaped pattern.

\begin{figure}[H]
   \begin{adjustwidth}{-\extralength}{0cm}
   \centering
 \includegraphics[width=8cm, angle=0]{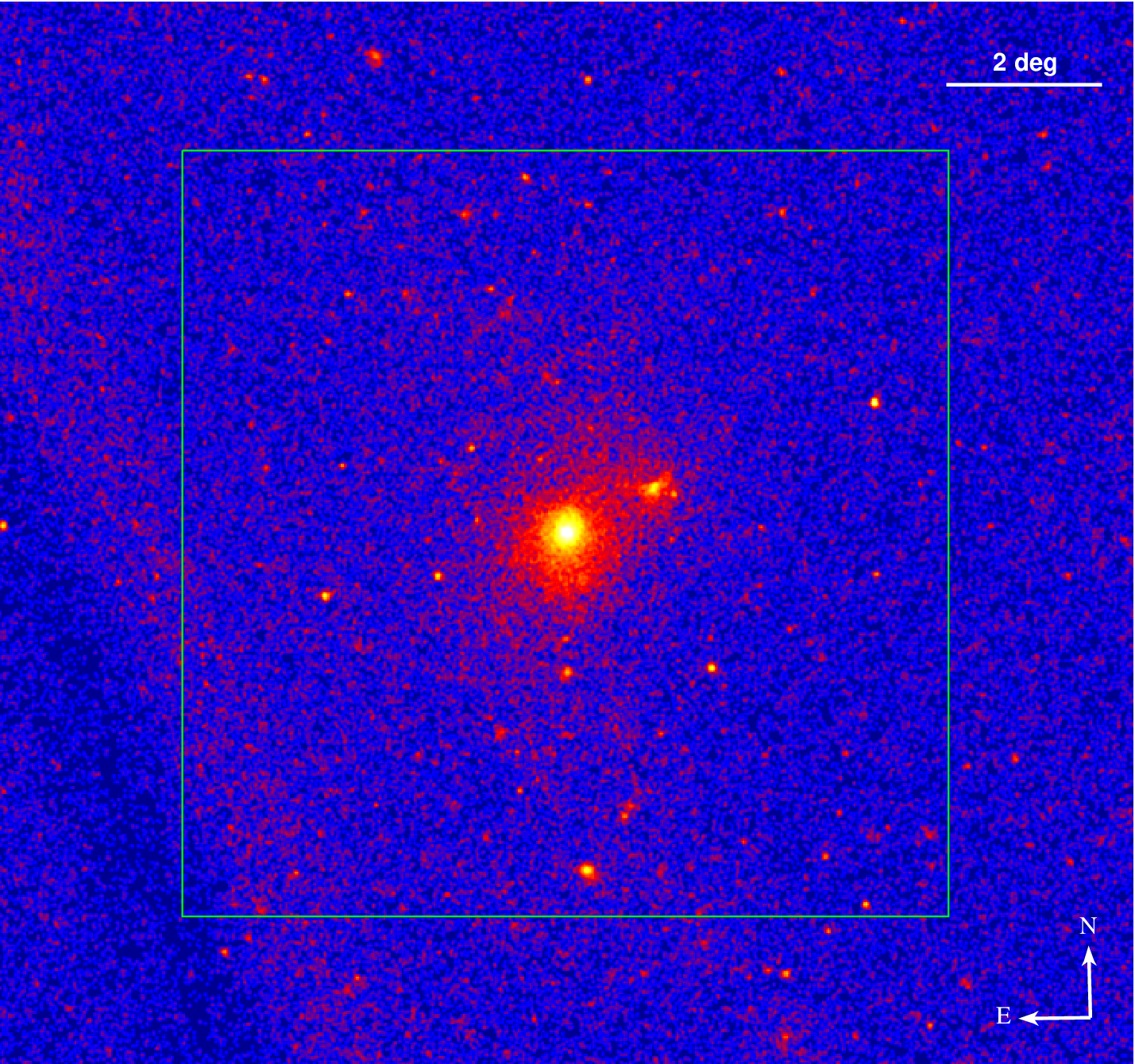}
    \hspace{0.3cm}
    \includegraphics[width=8cm, angle=0]{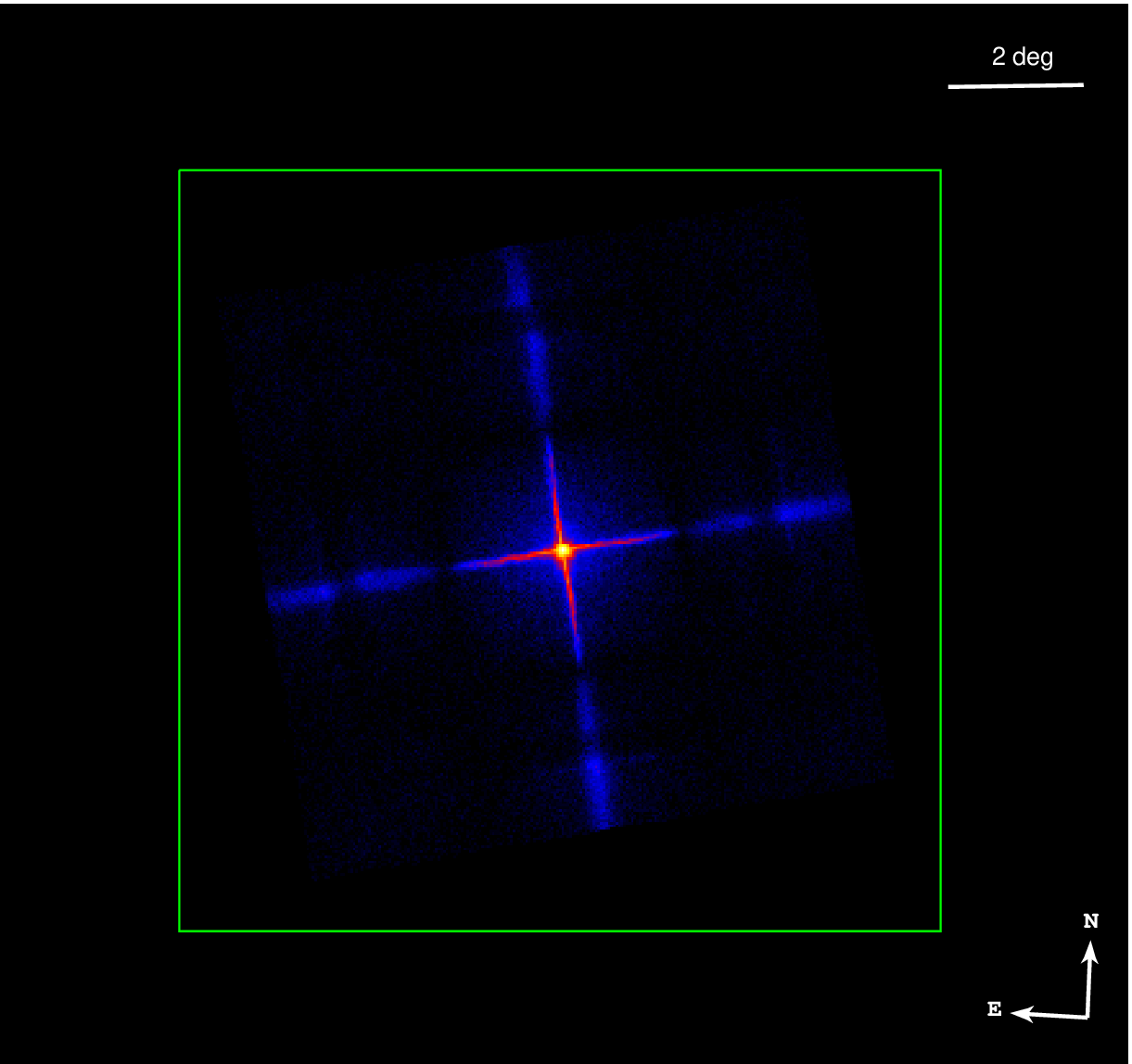}
\end{adjustwidth}
	\caption{(\textbf{Left}) ROSAT PSPC image of the Virgo Cluster in the 0.5--2.0 keV band. The center of the green box ($10^{\circ} \times 10^{\circ}$) is the location of the cluster's central AGN, M87. (\textbf{Right}) The X-ray image of Sco X-1 captured by LEIA that is regarded as the ideal PSF image of the observed source.}
	\label{Fig5}
\end{figure}

To simulate the imaging of ROSAT images under the influence of PSF, it is necessary to first generate a PSF model, which can be an ideal model or modeled based on the actual PSF data of LEIA. Then, using the ROSAT observation image as the source model, the source model is convolved with the PSF model to obtain the blurred PSF imaging result. LEIA observed Sco X-1, the brightest known persistent X-ray source \citep{Giacconi+etal+1962}, which is in great agreement with the PSF measured during on-ground calibration \citep{Zhang+etal+2022} (shown in the right panel of Figure ~\ref{Fig5}). Therefore, the X-ray image of Sco X-1 can be treated as an ideal PSF image for LEIA. The left panel in Figure ~\ref{Fig5} shows the ROSAT PSPC image of the Virgo Cluster in the 0.5--2.0 keV band. After rotating the PSF image to the average characteristics of the Virgo observations, we convolved it with the ROSAT image to simulate LEIA observations of the Virgo Cluster under the influence of the PSF effect. The FoV of the simulated image is $10^{\circ} \times 10^{\circ}$. 

\begin{figure}[H]
   \begin{adjustwidth}{-\extralength}{0cm}
   \centering
 \includegraphics[width=8.15cm, angle=0]{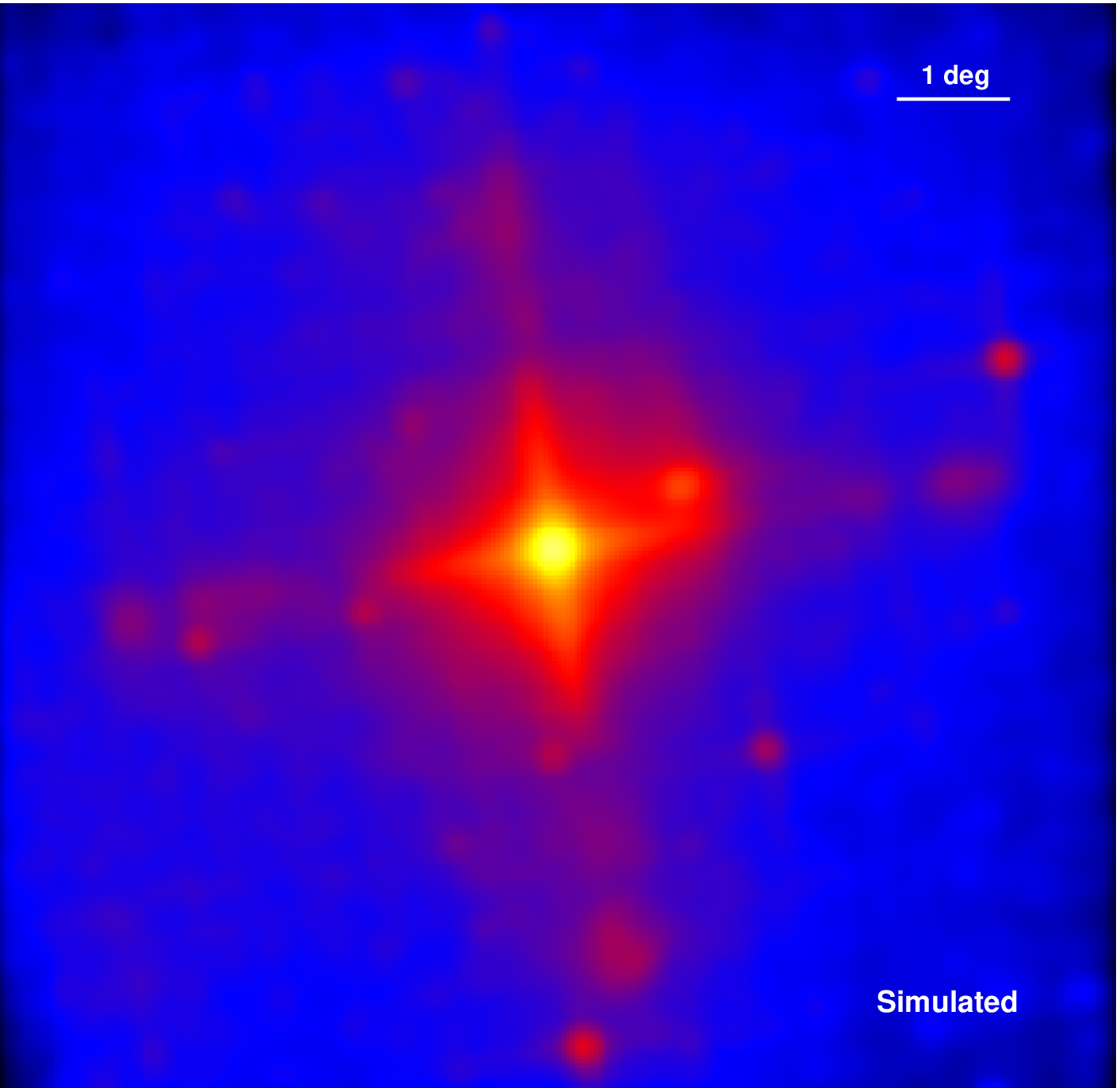}
    \hspace{0.3cm}
    \includegraphics[width=8.15cm, angle=0]{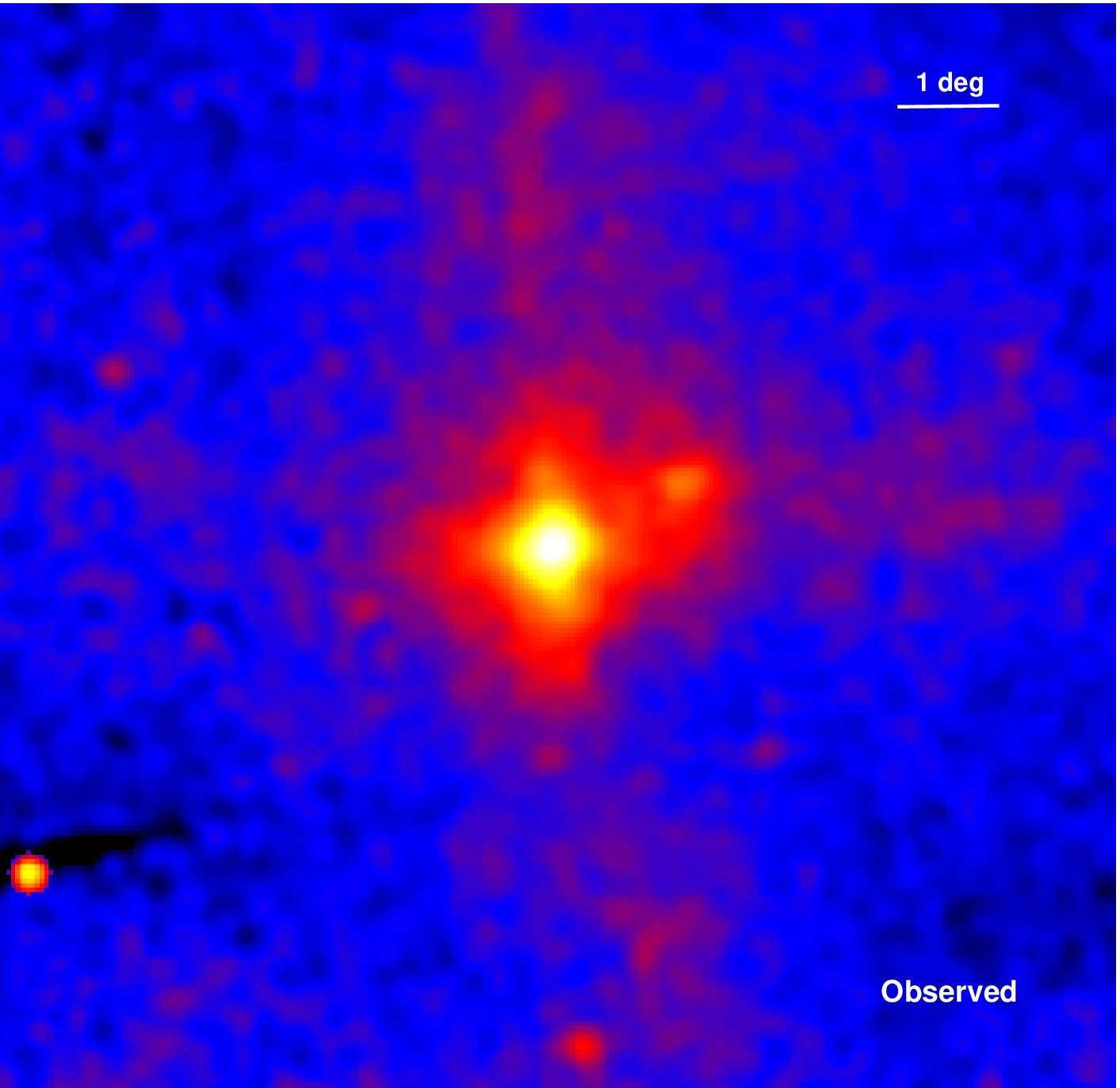}
\end{adjustwidth}
	\caption{(\textbf{Left}) The smoothed simulated observation of the source in the 0.5--2.0 keV band, covering the field of view of the green box in the left panel of Figure~\ref{Fig5} . (\textbf{Right}) The smoothed X-ray image of the central region obtained by LEIA with the same energy band and FoV as the left panel. }
	\label{Fig6}
\end{figure}

As shown in the left panel of Figure~\ref{Fig6}, there is an obvious arm structure extending outward from the brightest center of the simulated image. Comparing the simulated image with the observed one (the right panel of Figure~\ref{Fig6}), it can be seen that their extension structures are consistent. Therefore, we infer that there are no additional large-scale structures in the Virgo Cluster according to the LEIA observations.

\section{Conclusions}
\label{conclusions}
In this work, we present wide-field ($18.6^{\circ} \times 18.6^{\circ}$) X-ray observations of the Virgo Cluster obtained using LEIA and compare these observations with ROSAT images. We also study the LEIA spectrum. The main results are as follows:

\begin{enumerate}
  \item The images match the ROSAT PSPC images. Galaxies, including M87, M86, and M49, can be clearly seen from the merged LEIA image in the 0.5--4.5 keV band.
  \item  \textls[-15]{The LEIA images show an arm structure caused by the PSF effect of the LEIA instrument.}
  \item  We extracted the LEIA spectrum and determined the average temperature of the FoV ($2.1^{+0.3}_{-0.1}$ keV ) and found it to be consistent with the overall temperature ($2.4^{+0.3}_{-0.2}$ keV) within a 1$^{\circ}$ radius of M87, as measured by ROSAT, and the average temperature of ($\sim$2.3 keV) obtained by XMM-Newton.
  \item Above 1.5 keV, the spectrum is dominated by the background, with the image primarily showing the central point source and its PSF effects.
\end{enumerate}

This is the first time that the ability of the lobster-eye focusing telescope to generate images and spectra has been verified; however, the images are significantly affected by PSF. Future technological progress in eliminating the impact of PSF will greatly enhance the imaging ability of this type of telescope.

\vspace{6pt}

\authorcontributions{Conceptualization, Shu-Mei Jia and Heng Yu; Methodology, Wen-Cheng Feng, Cheng-Kui Li, Yu-Lin Cheng and Shan-Shan Weng; Software, Wen-Cheng Feng, Cheng-Kui Li and Yu-Lin Cheng; Validation, Shu-Mei Jia, Hai-Hui Zhao and Yong Chen; Formal analysis, Wen-Cheng Feng; Investigation, Heng Yu and Yong Chen; Resources, Wen-Cheng Feng, Hai-Wu Pan, Yuan Liu,  Zhi-Xing Ling and Chen Zhang; Data curation, Wen-Cheng Feng, Hai-Wu Pan and Yuan Liu; Writing – original draft, Wen-Cheng Feng; Writing – review \& editing, Heng Yu; Funding acquisition, Shu-Mei Jia, Hai-Hui Zhao and Shan-Shan Weng; Project administration, Shu-Mei Jia; Supervision, Shu-Mei Jia and Shan-Shan Weng; Visualization, Wen-Cheng Feng. }

\funding{This work was supported by the Strategic Pioneer Program on Space Science, Chinese Academy of Sciences (Grant No. XDA15310300), and the National Natural Science Foundation of China (Grant Nos. U2038104 and 11703014).
This research was carried out using Montage. It was funded by the National Science Foundation under Grant Number ACI-1440620 and was previously funded by the National Aeronautics and Space Administration's Earth Science Technology Office, Computation Technologies Project, under Cooperative Agreement Number NCC5-626 between NASA and the California Institute of Technology.}

\dataavailability{The data in this article has not been made public yet.  } 

\conflictsofinterest{The authors declare no conflict of interest. } 

\begin{adjustwidth}{-\extralength}{0cm}
\printendnotes[custom] 

\reftitle{References}

\PublishersNote{}
\end{adjustwidth}

\begin{thebibliography}{999}

\bibitem[{Angel}(1979)]{1979ApJ...233..364A}
{Angel}, J.R.P.
\newblock {Lobster eyes as X-ray telescopes.}
\newblock {\em  Astrophys. J.} {\bf 1979}, {\em 233},~364--373.
\newblock {\url{https://doi.org/10.1086/157397}}.

\bibitem[{Wilkins} et~al.(1989){Wilkins}, {Stevenson}, {Nugent}, {Chapman}, and
  {Steenstrup}]{1989RScI...60.1026W}
{Wilkins}, S.W.; {Stevenson}, A.W.; {Nugent}, K.A.; {Chapman}, H.;
  {Steenstrup}, S.
\newblock {On the concentration, focusing, and collimation of x-rays and
  neutrons using microchannel plates and configurations of holes}.
\newblock {\em Rev. Sci. Instrum.} {\bf 1989}, {\em
  60},~1026--1036.
\newblock {\url{https://doi.org/10.1063/1.1140312}}.

\bibitem[{Chapman} et~al.(1991){Chapman}, {Nugent}, and
  {Wilkins}]{1991RScI...62.1542C}
{Chapman}, H.N.; {Nugent}, K.A.; {Wilkins}, S.W.
\newblock {X-ray focusing using square channel-capillary arrays}.
\newblock {\em Rev. Sci. Instrum.} {\bf 1991}, {\em
  62},~1542--1561.
\newblock {\url{https://doi.org/10.1063/1.1142432}}.

\bibitem[Fraser et~al.(1992)Fraser, Lees, Pearson, Sims, and
  Roxburgh]{10.1117/12.51224}
Fraser, G.W.; Lees, J.E.; Pearson, J.F.; Sims, M.R.; Roxburgh, K.
\newblock {X-ray focusing using microchannel plates}.
\newblock In \emph{Proceedings of the Multilayer and Grazing Incidence X-Ray/EUV
  Optics}; Hoover, R.B., Ed.; International Society for Optics and Photonics: San Diego, CA, USA,
 1992; Volume 1546, pp. 41--52.
\newblock {\url{https://doi.org/10.1117/12.51224}}.

\bibitem[{Fraser} et~al.(1993){Fraser}, {Brunton}, {Lees}, {Pearson}, and
  {Feller}]{1993NIMPA.324..404F}
{Fraser}, G.W.; {Brunton}, A.N.; {Lees}, J.E.; {Pearson}, J.F.; {Feller}, W.B.
\newblock {X-ray focusing using square-pore microchannel plates First
  observation of cruxiform image structure}.
\newblock {\em Nucl. Instrum. Methods Phys. Res. A} {\bf
  1993}, {\em 324},~404--407.
\newblock {\url{https://doi.org/10.1016/0168-9002(93)91003-6}}.

\bibitem[Zhao et~al.(2014)Zhao, Zhang, Yuan, Willingale, Ling, Feng, Li, Ji,
  Wang, and Zhang]{10.1117/12.2055434}
Zhao, D.; Zhang, C.; Yuan, W.; Willingale, R.; Ling, Z.; Feng, H.; Li, H.; Ji,
  J.; Wang, W.; Zhang, S.
\newblock {Ray tracing simulations for the wide-field x-ray telescope of the
  Einstein Probe mission based on Geant4 and XRTG4}.
\newblock In \emph{Proceedings of the Space Telescopes and Instrumentation 2014:
  Ultraviolet to Gamma Ray}; Takahashi, T., den Herder, J.W.A., Bautz, M., Eds.; 
  International Society for Optics and Photonics: San Diego, CA, USA,
    2014; Volume 9144, p.
  91444E.
\newblock {\url{https://doi.org/10.1117/12.2055434}}.

\bibitem[{Hudec} et~al.(2017){Hudec}, {Sveda}, {P{\'\i}na}, {Inneman},
  {Semencova}, and {Skulinova}]{2017SPIE10567E..19H}
{Hudec}, R.; {Sveda}, L.; {P{\'\i}na}, L.; {Inneman}, A.; {Semencova}, V.;
  {Skulinova}, M.
\newblock {LOBSTER: new space x-ray telescopes}.
\newblock In \emph{Proceedings of the Society of Photo-Optical Instrumentation
  Engineers (SPIE) Conference Series};  San Diego, CA, USA, 2017; Volume 10567.
\newblock {\url{https://doi.org/10.1117/12.2308126}}.

\bibitem[Priedhorsky et~al.(1996)Priedhorsky, Peele, and
  Nugent]{10.1093/mnras/279.3.733}
Priedhorsky, W.C.; Peele, A.G.; Nugent, K.A.
\newblock {An X-ray all-sky monitor with extraordinary sensitivity}.
\newblock {\em Mon. Not. R. Astron. Soc.} {\bf 1996},
  {\em 279},~733--750. 
\newblock {\url{https://doi.org/10.1093/mnras/279.3.733}}.

\bibitem[Fraser et~al.(2002)Fraser, Brunton, Bannister, Pearson, Ward,
  Stevenson, Watson, Warwick, Whitehead, O'Brian, White, Jahoda, Black, Hunter,
  Deines-Jones, Priedhorsky, Brumby, Borozdin, Vestrand, Fabian, Nugent, Peele,
  Irving, Price, Eckersley, Renouf, Smith, Parmar, McHardy, Uttley, and
  Lawrence]{10.1117/12.454217}
Fraser, G.W.; Brunton, A.N.; Bannister, N.P.; Pearson, J.F.; Ward, M.;
  Stevenson, T.J.; Watson, D.J.; Warwick, B.; Whitehead, S.; O'Brian, P.;
  et~al.
\newblock {LOBSTER-ISS: an imaging x-ray all-sky monitor for the International
  Space Station}.
\newblock In \emph{Proceedings of the X-Ray and Gamma-Ray Instrumentation for
  Astronomy XII}; Flanagan, K.A., Siegmund, O.H.W., Eds.; International Society
  for Optics and Photonics: San Diego, CA, USA,  2002;  Volume 4497, pp. 115--126.
\newblock {\url{https://doi.org/10.1117/12.454217}}.

\bibitem[{Yuan} et~al.(2016){Yuan}, {Amati}, {Cannizzo}, {Cordier}, {Gehrels},
  {Ghirlanda}, {G{\"o}tz}, {Produit}, {Qiu}, {Sun}, {Tanvir}, {Wei}, and
  {Zhang}]{2016SSRv..202..235Y}
{Yuan}, W.; {Amati}, L.; {Cannizzo}, J.K.; {Cordier}, B.; {Gehrels}, N.;
  {Ghirlanda}, G.; {G{\"o}tz}, D.; {Produit}, N.; {Qiu}, Y.; {Sun}, J.;  et~al.
\newblock {Perspectives on Gamma-Ray Burst Physics and Cosmology with Next
  Generation Facilities}.
\newblock {\em Space Sci. Rev.} {\bf 2016}, {\em 202},~235--277.
\newblock {\url{https://doi.org/10.1007/s11214-016-0274-z}}.

\bibitem[O'Brien et~al.(2020)O'Brien, Hutchinson, Lerman, Feldman, McHugh,
  Lodge, Willingale, Beardmore, Speight, and Drumm]{10.1117/12.2561301}
O'Brien, P.; Hutchinson, I.; Lerman, H.; Feldman, C.; McHugh, M.; Lodge, A.;
  Willingale, R.; Beardmore, A.; Speight, R.; Drumm, P.
\newblock {The soft x-ray imager on THESEUS: the transient high energy survey
  and early universe surveyor}.
\newblock In \emph{Proceedings of the Space Telescopes and Instrumentation 2020:
  Ultraviolet to Gamma Ray}; den Herder, J.W.A., Nikzad, S., Nakazawa, K., Eds.; 
  International Society for Optics and Photonics: San Diego, CA, USA,
    2020; Volume 11444, p.
  114442L.
\newblock {\url{https://doi.org/10.1117/12.2561301}}.

\bibitem[{Bunce} et~al.(2020){Bunce}, {Martindale}, {Lindsay}, {Muinonen},
  {Rothery}, {Pearson}, {McDonnell}, {Thomas}, {Thornhill}, {Tikkanen},
  {Feldman}, {Huovelin}, {Korpela}, {Esko}, {Lehtolainen}, {Treis}, {Majewski},
  {Hilchenbach}, {V{\"a}is{\"a}nen}, {Luttinen}, {Kohout}, {Penttil{\"a}},
  {Bridges}, {Joy}, {Alcacera-Gil}, {Alibert}, {Anand}, {Bannister},
  {Barcelo-Garcia}, {Bicknell}, {Blake}, {Bland}, {Butcher}, {Cheney},
  {Christensen}, {Crawford}, {Crawford}, {Dennerl}, {Dougherty}, {Drumm},
  {Fairbend}, {Genzer}, {Grande}, {Hall}, {Hodnett}, {Houghton}, {Imber},
  {Kallio}, {Lara}, {Balado Margeli}, {Mas-Hesse}, {Maurice}, {Milan},
  {Millington-Hotze}, {Nenonen}, {Nittler}, {Okada}, {Orm{\"o}},
  {Perez-Mercader}, {Poyner}, {Robert}, {Ross}, {Pajas-Sanz}, {Schyns},
  {Seguy}, {Str{\"u}der}, {Vaudon}, {Viceira-Mart{\'\i}n}, {Williams},
  {Willingale}, and {Yeoman}]{2020SSRv..216..126B}
{Bunce}, E.J.; {Martindale}, A.; {Lindsay}, S.; {Muinonen}, K.; {Rothery},
  D.A.; {Pearson}, J.; {McDonnell}, I.; {Thomas}, C.; {Thornhill}, J.;
  {Tikkanen}, T.;  et~al.
\newblock {The BepiColombo Mercury Imaging X-Ray Spectrometer: Science Goals,
  Instrument Performance and Operations}.
\newblock {\em Space Sci. Rev.} {\bf 2020}, {\em 216},~126.
\newblock {\url{https://doi.org/10.1007/s11214-020-00750-2}}.

\bibitem[G{\"o}tz et~al.(2016)G{\"o}tz, Meuris, Pinsard, Doumayrou, Tourrette,
  Osborne, Willingale, Sykes, Pearson, Duigou, and Mercier]{10.1117/12.2232484}
G{\"o}tz, D.; Meuris, A.; Pinsard, F.; Doumayrou, E.; Tourrette, T.; Osborne,
  J.P.; Willingale, R.; Sykes, J.M.; Pearson, J.F.; Duigou, J.M.L.;  et~al.
\newblock {The microchannel x-ray telescope status}.
\newblock In \emph{Proceedings of the Space Telescopes and Instrumentation 2016:
  Ultraviolet to Gamma Ray}; den Herder, J.W.A., Takahashi, T., Bautz, M., Eds.; 
  International Society for Optics and Photonics: San Diego, CA, USA,
    2016; Volume 9905, p.
  99054L.
\newblock {\url{https://doi.org/10.1117/12.2232484}}.

\bibitem[Feldman et~al.(2022)Feldman, Willingale, Pearson, Butcher, Peterson,
  Crawford, Houghton, Speight, Lodge, Bicknell, Osborne, O'Brien, Bradshaw,
  Burwitz, Hartner, Langmeier, Mueller, Rukdee, Schmidt, G{\"o}tz, Mercier,
  Duigou, Gonzalez, Schyns, Roudot, Fairbend, and Seguy]{10.1117/12.2628635}
Feldman, C.; Willingale, R.; Pearson, J.; Butcher, G.; Peterson, P.; Crawford,
  T.; Houghton, P.; Speight, R.; Lodge, A.; Bicknell, C.;  et~al.
\newblock {Calibration of the flight model lobster eye optic for SVOM}.
\newblock In \emph{Proceedings of the Space Telescopes and Instrumentation 2022:
  Ultraviolet to Gamma Ray}; den Herder, J.W.A., Nikzad, S., Nakazawa, K., Eds.; 
  International Society for Optics and Photonics: San Diego, CA, USA,
   2022; Volume 12181, p.
  121811P.
\newblock {\url{https://doi.org/10.1117/12.2628635}}.

\bibitem[{Sembay} et~al.(2016){Sembay}, {Branduardi-Raymont}, {Drumm},
  {Escoubet}, {Genov}, {Gow}, {Hall}, {Holland}, {Hudec}, {Mas-Hesse},
  {Kennedy}, {Kuntz}, {Nakamura}, {Ostgaard}, {Ottensamer}, {Raab}, {Read},
  {Rebuffat}, {Romstedt}, {Schyns}, {Sibeck}, {Srp}, {Steller}, {Sun}, {Sykes},
  {Thornhill}, {Walsh}, {Walton}, {Wang}, {Wei}, {Wielders}, and
  {Whittaker}]{2016AGUFMSM44A..04S}
{Sembay}, S.; {Branduardi-Raymont}, G.; {Drumm}, P.; {Escoubet}, C.P.; {Genov},
  G.; {Gow}, J.; {Hall}, D.; {Holland}, A.; {Hudec}, R.; {Mas-Hesse}, J.M.;
  et~al.
\newblock {The Soft X-ray Imager (SXI) on the SMILE Mission}.
\newblock In Proceedings of the AGU Fall Meeting Abstracts, San Francisco, CA, USA, 12--16 December
 2016; p.
  SM44A-04.

\bibitem[Ling et~al.(2023)Ling, Sun, Zhang, Sun, Jin, Zhang, Zhang, Chang,
  Chen, Chen, Cheng, Fu, Han, Li, Li, Li, Li, Liu, Lv, Ma, Tang, Wang, Xie,
  Xue, Yan, Zhang, Bao, Cai, Cheng, Cui, Dai, Fan, Hu, Hu, Huang, Jia, Jin, Li,
  Li, Liu, Liu, Liu, Pan, Qiu, Sugizaki, Sun, Wang, Wang, Wu, Xu, Xu, Yang,
  Yang, Zhang, Zhang, Zhang, Zhang, Zhao, Cong, Jiang, Li, Qiu, Sun, Su, Wang,
  Wu, Xu, Yang, Zhang, Zhang, Zhang, Zhu, Ban, Bi, Cai, Chen, Chen, Chen, Cui,
  Duan, Feng, Gao, He, He, Huang, Li, Li, Li, Li, Liu, Liu, Liu, Liu, Meng,
  Shi, Sun, Wang, Wang, Wu, Xu, Yang, Yang, Yu, Zhang, Zhang, Zhang, Zhang,
  Zhou, Zhu, Cheng, Qin, Wang, Wang, Bai, Gao, Ji, Liu, Ma, Shi, Su, Tan, Tong,
  Xu, Xue, Xue, and Yuan]{Ling_2023}
Ling, Z.X.; Sun, X.J.; Zhang, C.; Sun, S.L.; Jin, G.; Zhang, S.N.; Zhang, X.F.;
  Chang, J.B.; Chen, F.S.; Chen, Y.F.;  et~al.
\newblock The Lobster Eye Imager for Astronomy Onboard the SATech-01 Satellite.
\newblock {\em Res. Astron. Astrophys.} {\bf 2023}, {\em
  23},~095007.
\newblock {\url{https://doi.org/10.1088/1674-4527/acd593}}.

\bibitem[{Zhang} et~al.(2022){Zhang}, {Ling}, {Sun}, {Sun}, {Liu}, {Li}, {Xue},
  {Chen}, {Dai}, {Jia}, {Liu}, {Zhang}, {Zhang}, {Zhang}, {Chen}, {Cheng},
  {Fu}, {Han}, {Li}, {Li}, {Li}, {Liu}, {Ma}, {Tang}, {Wang}, {Xie}, {Yan},
  {Zhang}, {Jiang}, {Jin}, {Li}, {Qiu}, {Su}, {Sun}, {Xu}, {Zhang}, {Zhang},
  {Zhang}, {Bi}, {Cai}, {He}, {Liu}, {Zhu}, {Cheng}, {Cui}, {Fan}, {Hu},
  {Huang}, {Jin}, {Li}, {Pan}, {Wang}, {Xu}, {Yang}, {Zhang}, {Zhang}, {Zhang},
  {Zhao}, {Bai}, {Ji}, {Liu}, {Ma}, {Su}, {Tong}, {Wang}, {Zhao}, {Feldman},
  {O'Brien}, {Osborne}, {Willingale}, {Burwitz}, {Hartner}, {Langmeier},
  {M{\"u}ller}, {Rukdee}, {Schmidt}, {Kuulkers}, and {Yuan}]{Zhang+etal+2022}
{Zhang}, C.; {Ling}, Z.X.; {Sun}, X.J.; {Sun}, S.L.; {Liu}, Y.; {Li}, Z.D.;
  {Xue}, Y.L.; {Chen}, Y.F.; {Dai}, Y.F.; {Jia}, Z.Q.;  et~al.
\newblock {First Wide Field-of-view X-Ray Observations by a Lobster-eye
  Focusing Telescope in Orbit}.
\newblock {\em  Astrophys. J. Lett.} {\bf 2022}, {\em 941},~L2. 
\newblock {\url{https://doi.org/10.3847/2041-8213/aca32f}}.

\bibitem[{Tonry} et~al.(2001){Tonry}, {Dressler}, {Blakeslee}, {Ajhar},
  {Fletcher}, {Luppino}, {Metzger}, and {Moore}]{2001ApJ...546..681T}
{Tonry}, J.L.; {Dressler}, A.; {Blakeslee}, J.P.; {Ajhar}, E.A.; {Fletcher},
  A.B.; {Luppino}, G.A.; {Metzger}, M.R.; {Moore}, C.B.
\newblock {The SBF Survey of Galaxy Distances. IV. SBF Magnitudes, Colors, and
  Distances}.
\newblock {\em  Astrophys. J.} {\bf 2001}, {\em 546},~681--693. 
\newblock {\url{https://doi.org/10.1086/318301}}.

\bibitem[{B{\"o}hringer} et~al.(1994){B{\"o}hringer}, {Briel}, {Schwarz},
  {Voges}, {Hartner}, and {Tr{\"u}mper}]{Bohringer+etal+1994}
{B{\"o}hringer}, H.; {Briel}, U.G.; {Schwarz}, R.A.; {Voges}, W.; {Hartner},
  G.; {Tr{\"u}mper}, J.
\newblock {The structure of the Virgo cluster of galaxies from Rosat X-ray
  images}.
\newblock {\em Nature} {\bf 1994}, {\em 368},~828--831.
\newblock {\url{https://doi.org/10.1038/368828a0}}.

\bibitem[{Truemper}(1992)]{Truemper+1992}
{Truemper}, J.
\newblock {ROSAT: A New Look at the X-ray Sky}.
\newblock {\em  Q. J. R. Astron. Soc.} {\bf
  1992}, {\em 33},~165.

\bibitem[{Shibata} et~al.(2001){Shibata}, {Matsushita}, {Yamasaki}, {Ohashi},
  {Ishida}, {Kikuchi}, {B{\"o}hringer}, and {Matsumoto}]{Shibata+etal+2001}
{Shibata}, R.; {Matsushita}, K.; {Yamasaki}, N.Y.; {Ohashi}, T.; {Ishida}, M.;
  {Kikuchi}, K.; {B{\"o}hringer}, H.; {Matsumoto}, H.
\newblock {Temperature Map of the Virgo Cluster of Galaxies Observed with
  ASCA}.
\newblock {\em  Astrophys. J.} {\bf 2001}, {\em 549},~228--243. 
\newblock {\url{https://doi.org/10.1086/319075}}.

\bibitem[{Urban} et~al.(2011){Urban}, {Werner}, {Simionescu}, {Allen}, and
  {B{\"o}hringer}]{Urban+etal+2011}
{Urban}, O.; {Werner}, N.; {Simionescu}, A.; {Allen}, S.W.; {B{\"o}hringer}, H.
\newblock {X-ray spectroscopy of the Virgo Cluster out to the virial radius}.
\newblock {\em Mon. Not. R. Astron. Soc.} {\bf 2011}, {\em 414},~2101--2111. 
\newblock {\url{https://doi.org/10.1111/j.1365-2966.2011.18526.x}}.

\bibitem[{Cappellari} et~al.(2011){Cappellari}, {Emsellem}, {Krajnovi{\'c}},
  {McDermid}, {Scott}, {Verdoes Kleijn}, {Young}, {Alatalo}, {Bacon}, {Blitz},
  {Bois}, {Bournaud}, {Bureau}, {Davies}, {Davis}, {de Zeeuw}, {Duc},
  {Khochfar}, {Kuntschner}, {Lablanche}, {Morganti}, {Naab}, {Oosterloo},
  {Sarzi}, {Serra}, and {Weijmans}]{Cappellari+etal+2011}
{Cappellari}, M.; {Emsellem}, E.; {Krajnovi{\'c}}, D.; {McDermid}, R.M.;
  {Scott}, N.; {Verdoes Kleijn}, G.A.; {Young}, L.M.; {Alatalo}, K.; {Bacon},
  R.; {Blitz}, L.;  et~al.
\newblock {The ATLAS$^{3D}$ project---I. A volume-limited sample of 260 nearby
  early-type galaxies: science goals and selection criteria}.
\newblock {\em Mon. Not. R. Astron. Soc.} {\bf 2011}, {\em 413},~813--836. 
\newblock {\url{https://doi.org/10.1111/j.1365-2966.2010.18174.x}}.

\bibitem[{Liu} et~al.(2024){Liu}, {Sun}, {Xu}, {Svinkin}, {Delaunay}, {Tanvir},
  {Gao}, {Zhang}, {Chen}, {Wu}, {Zhang}, {Yuan}, {An}, {Bruni}, {Frederiks},
  {Ghirlanda}, {Hu}, {Li}, {Li}, {Li}, {Malesani}, {Piro}, {Raman}, {Ricci},
  {Troja}, {Vergani}, {Wu}, {Yang}, {Zhang}, {Zhu}, {de Ugarte Postigo},
  {Demin}, {Dobie}, {Fan}, {Fu}, {Fynbo}, {Geng}, {Gianfagna}, {Hu}, {Huang},
  {Jiang}, {Jonker}, {Julakanti}, {Kennea}, {Kokomov}, {Kuulkers}, {Lei},
  {Leung}, {Levan}, {Li}, {Li}, {Littlefair}, {Liu}, {Lysenko}, {Ma},
  {Martin-Carrillo}, {O'Brien}, {Parsotan}, {Quirola-Vasquez}, {Ridnaia},
  {Ronchini}, {Rossi}, {Mata-Sanchez}, {Schneider}, {Shen}, {Thakur},
  {Tohuvavohu}, {Torres}, {Tsvetkova}, {Ulanov}, {Wei}, {Xiao}, {Yin}, {Bai},
  {Burwitz}, {Cai}, {Chen}, {Chen}, {Chen}, {Chen}, {Chen}, {Chen}, {Cheng},
  {Cui}, {Cui}, {Dai}, {Dai}, {Eder}, {Fan}, {Feldman}, {Feng}, {Feng},
  {Friedrich}, {Gao}, {Guan}, {Han}, {Han}, {Hou}, {Hu}, {Hu}, {Huang}, {Huo},
  {Hutchinson}, {Ji}, {Jia}, {Jia}, {Jiang}, {Jin}, {Jin}, {Jin}, {Keereman},
  {Lerman}, {Li}, {Li}, {Li}, {Li}, {Li}, {Lian}, {Liang}, {Ling}, {Liu},
  {Liu}, {Liu}, {Liu}, {Liu}, {Lu}, {LU}, {Luo}, {Ma}, {Ma}, {Mao}, {Mao},
  {McHugh}, {Meidinger}, {Nandra}, {Osborne}, {Pan}, {Pan}, {Ravasio}, {Rau},
  {Rea}, {Rehman}, {Sanders}, {Santovincenzo}, {Song}, {Su}, {Sun}, {Sun},
  {Sun}, {Tan}, {Tang}, {Tao}, {Tong}, {Wang}, {Wang}, {Wang}, {Wang}, {Wang},
  {Wang}, {Wang}, {Wang}, {Wei}, {Willingale}, {Xiong}, {Xu}, {Xu}, {Xu}, {Xu},
  {Xu}, {Xue}, {Xue}, {Yan}, {Yang}, {Yang}, {Yang}, {Yang}, {Yu}, {Zhang},
  {Zhang}, {Zhang}, {Zhang}, {Zhang}, {Zhang}, {Zhang}, {Zhang}, {Zhang},
  {Zhao}, {Zhao}, {Zhao}, {Zhao}, {Zhou}, {Zhou}, {Zhu}, {Zhu}, and
  {Zuo}]{2024arXiv240416425L}
{Liu}, Y.; {Sun}, H.; {Xu}, D.; {Svinkin}, D.S.; {Delaunay}, J.; {Tanvir},
  N.R.; {Gao}, H.; {Zhang}, C.; {Chen}, Y.; {Wu}, X.F.;  et~al.
\newblock {Soft X-ray prompt emission from a high-redshift gamma-ray burst
  EP240315a}.
\newblock {\em arXiv} {\bf 2024}, arXiv:2404.16425. 
\newblock {\url{https://doi.org/10.48550/arXiv.2404.16425}}.

\bibitem[{Vollmer} et~al.(2013){Vollmer}, {Soida}, {Beck}, {Chung}, {Urbanik},
  {Chy{\.z}y}, {Otmianowska-Mazur}, and {Kenney}]{Vollmer+etal+2013}
{Vollmer}, B.; {Soida}, M.; {Beck}, R.; {Chung}, A.; {Urbanik}, M.;
  {Chy{\.z}y}, K.T.; {Otmianowska-Mazur}, K.; {Kenney}, J.D.P.
\newblock {Large-scale radio continuum properties of 19 Virgo cluster galaxies.
  The influence of tidal interactions, ram pressure stripping, and accreting
  gas envelopes}.
\newblock {\em Astron. Astrophys.} {\bf 2013}, {\em 553},~A116. 
\newblock {\url{https://doi.org/10.1051/0004-6361/201321163}}.

\bibitem[{White} et~al.(1991){White}, {Fabian}, {Forman}, {Jones}, and
  {Stern}]{White+etal+1991}
{White}, D.A.; {Fabian}, A.C.; {Forman}, W.; {Jones}, C.; {Stern}, C.
\newblock {Ram-Pressure Stripping of the Multiphase Interstellar Medium of the
  Virgo Cluster Elliptical Galaxy M86 (NGC 4406)}.
\newblock {\em Astrophys. J.} {\bf 1991}, {\em 375},~35.
\newblock {\url{https://doi.org/10.1086/170167}}.

\bibitem[{Binggeli} et~al.(1987){Binggeli}, {Tammann}, and
  {Sandage}]{Binggeli+etal+1987}
{Binggeli}, B.; {Tammann}, G.A.; {Sandage}, A.
\newblock {Studies of the Virgo Cluster. VI. Morphological and Kinematical
  Structure of the Virgo Cluster}.
\newblock {\em Astron. J. } {\bf 1987}, {\em 94},~251.
\newblock {\url{https://doi.org/10.1086/114467}}.

\bibitem[{Simionescu} et~al.(2007){Simionescu}, {B{\"o}hringer}, {Br{\"u}ggen},
  and {Finoguenov}]{Simionescu+etal+2007}
{Simionescu}, A.; {B{\"o}hringer}, H.; {Br{\"u}ggen}, M.; {Finoguenov}, A.
\newblock {The gaseous atmosphere of M 87 seen with XMM-Newton}.
\newblock {\em Astron. Astrophys.} {\bf 2007}, {\em 465},~749--758.
\newblock {\url{https://doi.org/10.1051/0004-6361:20066650}}.

\bibitem[{Matsushita} et~al.(2002){Matsushita}, {Belsole}, {Finoguenov}, and
  {B{\"o}hringer}]{Matsushita+etal+2002}
{Matsushita}, K.; {Belsole}, E.; {Finoguenov}, A.; {B{\"o}hringer}, H.
\newblock {XMM-Newton observation of M 87. I. Single-phase temperature
  structure of intracluster medium}.
\newblock {\em Astron. Astrophys.} {\bf 2002}, {\em 386},~77--96. 
\newblock {\url{https://doi.org/10.1051/0004-6361:20020087}}.

\bibitem[{Molendi}(2002)]{Molendi+2002}
{Molendi}, S.
\newblock {On the Temperature Structure of M87}.
\newblock {\em  Astrophys. J.} {\bf 2002}, {\em 580},~815--823. 
\newblock {\url{https://doi.org/10.1086/343757}}.

\bibitem[{Kalberla} et~al.(2005){Kalberla}, {Burton}, {Hartmann}, {Arnal},
  {Bajaja}, {Morras}, and {P{\"o}ppel}]{Kalberla+etal+2005}
{Kalberla}, P.M.W.; {Burton}, W.B.; {Hartmann}, D.; {Arnal}, E.M.; {Bajaja},
  E.; {Morras}, R.; {P{\"o}ppel}, W.G.L.
\newblock {The Leiden/Argentine/Bonn (LAB) Survey of Galactic HI. Final data
  release of the combined LDS and IAR surveys with improved stray-radiation
  corrections}.
\newblock {\em Astron. Astrophys.} {\bf 2005}, {\em 440},~775--782. 
\newblock {\url{https://doi.org/10.1051/0004-6361:20041864}}.

\bibitem[{Zhao} et~al.(2017){Zhao}, {Zhang}, {Yuan}, {Zhang}, {Willingale}, and
  {Ling}]{2017ExA....43..267Z}
{Zhao}, D.; {Zhang}, C.; {Yuan}, W.; {Zhang}, S.; {Willingale}, R.; {Ling}, Z.
\newblock {Geant4 simulations of a wide-angle x-ray focusing telescope}.
\newblock {\em Exp. Astron.} {\bf 2017}, {\em 43},~267--283. 
\newblock {\url{https://doi.org/10.1007/s10686-017-9534-5}}.

\bibitem[{Arnaud} et~al.(2005){Arnaud}, {Pointecouteau}, and
  {Pratt}]{Arnaud+etal+2005}
{Arnaud}, M.; {Pointecouteau}, E.; {Pratt}, G.W.
\newblock {The structural and scaling properties of nearby galaxy clusters. II.
  The M-T relation}.
\newblock {\em Astron. Astrophys.} {\bf 2005}, {\em 441},~893--903. 
\newblock {\url{https://doi.org/10.1051/0004-6361:20052856}}.

\bibitem[{Giacconi} et~al.(1962){Giacconi}, {Gursky}, {Paolini}, and
  {Rossi}]{Giacconi+etal+1962}
{Giacconi}, R.; {Gursky}, H.; {Paolini}, F.R.; {Rossi}, B.B.
\newblock {Evidence for x Rays From Sources Outside the Solar System}.
\newblock {\em Phys. Rev. Lett.} {\bf 1962}, {\em 9},~439--443.
\newblock {\url{https://doi.org/10.1103/PhysRevLett.9.439}}.

\end{thebibliography}
\end{document}